\documentclass[aps,pra,twocolumn,superscriptaddress]{revtex4}
\usepackage{amsmath}
\usepackage{latexsym}
\usepackage{amssymb}
\usepackage{graphics,epstopdf}
\usepackage{graphicx}
\usepackage[colorlinks=true, citecolor=blue, urlcolor=blue ]{hyperref}
\usepackage{float}
\usepackage{graphicx}
\usepackage{amsfonts}

\begin{document}

\title{Position and momentum cannot both be lazy: Quantum reciprocity relation with Lipschitz constants}

\author{Mahasweta Pandit}
\affiliation{Harish-Chandra Research Institute, HBNI, Chhatnag Road, Jhunsi, Allahabad 211 019, India}
\affiliation{National Institute of Technology, Rourkela, Odisha 769 008, India}

\author{Anindita Bera}
\affiliation{Harish-Chandra Research Institute, HBNI, Chhatnag Road, Jhunsi, Allahabad 211 019, India}
\affiliation{Department of Applied Mathematics, University of Calcutta, 92 A.P.C. Road, Kolkata 700 009, India}

\author{Aditi Sen(De)}
\affiliation{Harish-Chandra Research Institute, HBNI, Chhatnag Road, Jhunsi, Allahabad 211 019, India}

\author{Ujjwal Sen}
\affiliation{Harish-Chandra Research Institute, HBNI, Chhatnag Road, Jhunsi, Allahabad 211 019, India}

\begin{abstract}
We propose a trade-off between 
the Lipschitz constants of the position and
momentum probability distributions for arbitrary quantum states. We refer to the trade-off as a quantum reciprocity relation. The Lipschitz constant of a function
may be considered to quantify the extent of fluctuations 
of that function, and  is in general  
independent of its spread. The spreads of the position and momentum distributions are used to obtain the celebrated Heisenberg quantum uncertainty relations. We find that the
 product of the Lipschitz constants of position and momentum probability distributions is bounded below by a number that is of the order of
the inverse square of the Planck's constant.

\end{abstract}
\maketitle


\section{Introduction}
\label{intro}

\emph{[Please also see the response to comment on this manuscript, appended at the end of the manuscript.]}

Uncertainty relations are considered to be among the pillars
of quantum mechanics, both in terms of understanding and  utility of the latter
\cite{busch2006,uncertainty-bunch,kennard1,robertson,heisenberg-1930,scrodinger-vai,schrodinger-translation,birula1,Deutsch,Partovi,new-birula,kraus,Maassen1,birula-dekh,Vicente,berta-nature,zozor,Branciard1,Werner-uff,Pati-prl}. And so, while they are
useful in finding the border between classical and quantum worlds,
they are also effective to provide security in quantum cryptography
\cite{q-crypto-and-uncertainty,gisinRMP,koashi2006}. One of the most well-known among these relations is the Heisenberg uncertainty relation \cite{uncertainty-bunch,kennard1} between position and momentum distributions of an arbitrary quantum state.
In its usual form, it is conveyed as a constraint on the product of
spreads, quantified by standard deviations, of the position and momentum
probability distributions of an arbitrary quantum system, say, moving on a
line. As a consequence, if the profile of the position probability distribution of a
quantum system is very sharp (i.e., of low spread), the momentum
distribution is very broad (i.e., of high spread), and vice-versa.  
This is in sharp contrast to the case of a
classical point particle, which can have very
well-defined values of both position and momentum.

We address here a complementary question: What if the position
probability distribution of a quantum system moving on a line is devoid of
any features (fluctuations), i.e., it is flat (uniform distribution)? As
can be easily seen, the corresponding momentum probability distribution,
obtained through the Fourier transform~\cite{fourier-transform}, has a
single large spike (similar to a Dirac delta function~\cite{book-operational-calculus,dirac-delta}).
It therefore appears that the fluctuations
of position and
momentum distributions satisfy a trade-off, which we term as a ``quantum reciprocity relation''. We quantify the
fluctuation of a probability distribution by the square root of its Lipschitz
constant \cite{lip1,lip2-check,rudin-book,lip3,amritava-realanalysis}, assuming that the same exists.
Uniform (probability) distribution on an infinite line and the Dirac delta function are
not mathematically well-defined. Approximating them by Gaussian
probability distributions, we find that the product of the fluctuations for
position and momentum distributions equals \((1/\hbar)\sqrt{2/e\pi}\).
We
find by examining several examples, including the excited states of the
quantum one-dimensional simple harmonic oscillator, and quantum states
corresponding to Cauchy-Lorentz and Student's $t$ as position distributions, that
this Gaussian lower bound is not violated. Haar uniform generation over the space of Hermite polynomials however provide a numerical lower bound of the product of the fluctuations approximating \(0.3/\hbar\). The classical limit of the reciprocity in terms of 
the Lipschitz constants is quite unlike the same of the Heisenberg uncertainty in terms of spreads. In particular, here, the classical case falls within the quantum mechanically accessible region. 
We emphasize that the bound that is being proposed has not yet been proven analytically, and is being currently based on numerics and specific
examples.

We arrange the paper in the following way. In Sec.~\ref{lipschitz-cont}, we provide a brief recapitulation of Lipschitz continuous
functions, and the corresponding Lipschitz constants. In Sec.~\ref{method}, we
recount a method of evaluating the Lipschitz constant for differentiable
functions. A discussion on evidence for a non-zero lower bound of the product of fluctuations of Lipschitz constants is presented in Sec.~\ref{exist}.
 The case of eigenstates of the simple harmonic oscillator is
considered in Sec.~\ref{excited}. 
In Sec.~\ref{other-prob}, we consider quantum systems on the line, for which the position distributions are Cauchy-Lorentz or Student's $t$-distributions. Next, in Sec.~\ref{hermite}, we
use the completeness of Hermite polynomials on the space of square-integrable functions to determine the minimum-reciprocity state. 
In Sec.~\ref{Javed-Habib-BJP-te-join-korechhe}, we try to interpret the reciprocity relation, and compare it with the usual Heisenberg uncertainty relation. It also contains a discussion on the classical limit of the proposed reciprocity relation.  
A conclusion is presented in Section~\ref{conclude}.

\section{Lipschitz continuity and Lipschitz constant}
\label{lipschitz-cont}
A mapping $f:\mathcal{D} \rightarrow \mathbb{C}$  from domain $\mathcal{D}~(\subseteq \mathbb{R}^n)$ into the set of complex numbers $\mathbb{C}$ is said to be Lipschitz continuous if there exists a positive real constant $\eta$  such that
\begin{equation}
|f(x_{1}) - f(x_{2})| \leq \eta |x_{1}-x_{2}|,~~\forall ~x_{1},x_{2}\in \mathcal{D},
\label{lipschitz_continuity}
\end{equation}
where the modulus on the right-hand-side of inequality (\ref{lipschitz_continuity}) denotes the Euclidean norm of its argument. An important point to note is that while $\eta$ can depend on the domain, $\mathcal{D}$, it is independent of $x_1$ and $x_2$. Lipschitz continuous functions are of course continuous functions, but may not be differentiable. We refer to the minimal $\eta$ that satisfies inequality (\ref{lipschitz_continuity}) as the Lipschitz constant (LC) of $f$ in $\mathcal{D}$.
The Lipschitz continuity and the Lipschitz constant are designed to account for, and measure, changes of function values relative to changes in the independent variables in the \emph{entire} domain $\mathcal{D}$. The Lipschitz constant is certainly dependent on the function $f$, and varies from being small for one function to large for another. If $\eta$ is small, then $f(x)$ can change only a little, with a small change of $x$, while if $\eta$ is large, then $f(x)$ may have a large variation for only a small change of $x$.  Along with cases for bounded domains, we will also be interested in cases where $\mathcal{D}$ is unbounded. In particular, for quantum systems moving in one dimension, we are interested in the functions $f:[x_1,x_2] \to \mathbb{C}$ as well as the functions $f:\mathbb{R} \to \mathbb{C}$.

\section{The Methodology}
\label{method}
We begin here by briefly recapitulating a procedure to find the Lipschitz constant for any differentiable function $f(x)$. For this purpose, we consider the mean value theorem~\cite{rudin-book}, which states that if $f(x)$ is defined and continuous in the closed interval $[x_1,x_2]$ and differentiable in the open interval $(x_1,x_2)$, then there is at least one number $z$ in $(x_1,x_2)$ such that 
\begin{equation}
(f(x_{2}) - f(x_{1}))/(x_{2} - x_{1})  = f'(z).
\label{mean_value_theorem}
\end{equation}
The supremum of $|f'(z)|$ in $(x_1,x_2)$ will be the Lipschitz constant for $f$ in $[x_1,x_2]$.
Note however that this method is certainly not the only method for evaluating the LC. In particular, this will not work in those cases where the function is Lipschitz continuous but not differentiable, in the relevant domain.

\section{Evidence of a lower bound}
\label{exist}
While the considerations can be readily generalized to higher dimensions, we will mostly be considering one-dimensional systems. Consider, therefore, a quantum system moving along the $x$-axis, and described by the quantum wave function $\psi(x)$ in co-ordinate representation, with the position probability distribution given by $f(x) = |\psi(x)|^{2}$. Let the LC for $f$ be denoted by $\eta_x$. 
The momentum representation of the same quantum system is obtained by the Fourier transformation of $\psi(x)$:
\begin{equation}
\phi(p) = \frac{1}{\sqrt{2 \pi \hbar}} \int_{- \infty}^{\infty} \psi(x) e^{-i x p/\hbar} dx.
\label{phi_p}
\end{equation}
The momentum probability distribution is then given by $g(p) = |\phi(p)|^{2}$, and let us denote the LC for $g$ as $\eta_p$.

We want to focus on the characteristics of the products of Lipschitz constants of position and momentum probability distributions for arbitrary  quantum states. In particular, we want to study the quantity 
\begin{equation} 
\tilde{\eta}_x\tilde{\eta}_p,
\label{tilde_etax_etap} 
\end{equation}
where $\tilde{\eta}_x = \sqrt{\eta_{x}}$ and $\tilde{\eta}_p = \sqrt{\eta_{p}}$. 

For any position probability distribution $f(x)$, if the Lipschitz constant is zero, then 
$|f(x_{1}) - f(x_{2})|=0$ for all $x_1$, $x_2$, lying in the relevant domain, which implies that $f$ is a constant, say, $f(x) = c^{2}$, with $c$ being real. This function is not normalizable, and hence cannot be a probability distribution in the strict sense. But, as is usual, we will interpret it as the relative density of the number of particles, e.g. in a scattering experiment. The corresponding momentum probability distribution is a Dirac delta function in momentum space. A Dirac delta function
\begin{equation}
\label{dirac1}
\delta(\xi-\xi_0),
\end{equation}
 as a function of $\xi$, is unbounded at $\xi=\xi_0$ and vanishes elsewhere, so that its LC diverges. In a similar way, a vanishing $\eta_p$ implies a diverging $\eta_x$. Therefore, the Lipschitz constants of position and momentum probability distributions of a single quantum system cannot vanish simultaneously. This provides evidence for a non-zero lower bound for $\tilde{\eta}_x \tilde{\eta}_p$. We will get back to this point when we consider the ground state of simple harmonic oscillator, and again when the Cauchy-Lorentz distribution is investigated.
\section{Eigenstates of the simple harmonic oscillator}
\label{excited}
We begin with a paradigmatic quantum system, viz. the one-dimensional simple harmonic oscillator (SHO). Below we discuss the behavior of the products of LCs of position and momentum distributions corresponding to energy eigenstates of the SHO. The Gaussian probability distribution, given by
\begin{equation}
f_{g} (x:\mu,\sigma^{2}) = \frac{1}{\sqrt{2 \pi \sigma^2}} e^{-\frac{(x -\mu)^{2}}{2\sigma^{2}}}, ~~x\in\mathbb{R},
\label{gaussian_gen_def}
\end{equation}
where $\mu$ and $\sigma$ are respectively the mean and standard deviation of the distribution, will be of relevance here.
%

Let us first consider the ground state of a quantum simple harmonic oscillator of mass $m$. It is well-known that the position probability distribution of the ground state of the SHO is a Gaussian distribution and is given by $f_0(x)=f_{g}(x:0, \frac{1}{2 \alpha})$, where $\alpha = \frac{m \omega}{\hbar}$, with $\omega$ being the natural oscillator frequency, and the potential is chosen to be centered at the origin.
Now, to find the LC in position space, we need to calculate the maximum value of the derivative of $f_0(x)$, and we obtain
\begin{equation}
\eta_x=\sqrt{\frac{2}{e \pi}} \frac{m \omega}{\hbar}.
\label{gaussian_eta_x}
\end{equation}
The Fourier transformation of a Gaussian distribution is also a Gaussian distribution, and the corresponding momentum probability distribution function for the ground state of SHO is $g_0(p)=f_{g}(p:0, \frac{\alpha \hbar^2}{2})$. By approaching in a similar way as stated above, we find the LC of the momentum probability distribution to be
\begin{equation}
\eta_p = \sqrt{\frac{2}{e \pi}} \frac{1}{m \omega\hbar}.
\label{gaussian_eta_p}
\end{equation}
Therefore, for the ground state of SHO, the product of fluctuations for position and momentum distributions is given by
\begin{equation}
\tilde{\eta}_x \tilde{\eta}_p = \frac{1}{\hbar} \sqrt{\frac{2}{e \pi}} \approx\frac{0.4839}{\hbar}.
\label{gaussian_sqrt_etax_etap}
\end{equation}
Note that the reciprocity product of fluctuations of position and momentum is of the order of $\frac{1}{\hbar}$, whereas in the traditional uncertainty relation, the uncertainty product in terms of spreads of the position and momentum distributions for arbitrary quantum states is of the order of $\hbar$, or higher.

Let us now consider the case when $\alpha\rightarrow0$. In this limit, $f_0(x)$ tends to a constant  function.
In this case, as mentioned earlier around (\ref{dirac1}) and as seen from Eq.~(\ref{gaussian_eta_x}), the value of $\eta_{x}$ converges, linearly, to zero with $\alpha$. The corresponding momentum probability distribution can be considered to tend to the Dirac delta function, and has its LC diverging as $\frac{1}{\alpha}$.
Therfore, the quantity $\tilde{\eta}_x \tilde{\eta}_p$ does not depend on the value of $\alpha$, 
even when $\alpha\to0$. Therefore, if we approximate the constant and Dirac delta functions respectively by  sequences of Gaussian distributions, $f_g(x:0,\frac{\alpha_n \hbar^2}{2})$ and $f_g(x:0,\frac{1}{2 \alpha_n})$, with $\alpha_n \to 0$ as $n\to\infty$,
then the evidence of a \emph{non-zero} lower bound on $\tilde{\eta}_x \tilde{\eta}_p$, for arbitrary quantum states, alluded to around Eq. (\ref{dirac1}) in Sec. \ref{exist}, turns into a proof.

\subsection{Excited states of SHO}
We now want to investigate the behavior of the reciprocity product of LCs of position and momentum distributions,    
when the SHO is in its higher excitation levels. The wave functions in coordinate representation, corresponding to the excited states of the one-dimensional SHO, are given by
\begin{equation}
\psi_{n} (x) = \big(\frac{\alpha}{\pi}\big)^\frac{1}{4} \frac{1}{\sqrt{2^n n!}} H_n(\sqrt{\alpha} x) e^{-\frac{\alpha  x^2}{2}},~n=1,2,3,\ldots.
\label{gexcited_psi_x}
\end{equation}
Here, the function $H_n$ is the Hermite polynomial of $n^{\text{th}}$ order, and can be expressed as
\begin{equation}
H_n(z)=(-1)^n e^{z^2} \frac{d^n}{dz^n}(e^{-z^2}).
\end{equation}
The corresponding probability distribution in the position space is $f_{n}(x) = |\psi_{n} (x)|^{2}$. Note that the probability distributions corresponding to the energy eigenstates of the SHO in  position as well as in momentum spaces are differentiable in the entire spaces.

Now let us consider the first excited state of the SHO, which is obtained by setting  $n=1$. Then, the LC of the position probability distribution is obtained by maximizing the derivative of $f_1(x)$, 
so that 
$\eta_x \approx 0.6626~\alpha$.
Similarly, using momentum representation, we find that the LC of the momentum probability distribution corresponding to the first excited state of the SHO is $\eta_p \approx 0.6626/{\alpha \hbar^2}$.
 Therefore the corresponding reciprocity product of the fluctuations for position and momentum distributions is given by
\begin{equation}
\tilde{\eta}_x \tilde{\eta}_p \approx \frac{0.6626}{\hbar},
\end{equation}
which, just like for the ground state, is also of the order of $\frac{1}{\hbar}$, but of a higher value.

\begin{center}
\begin{figure}[t]
\includegraphics[width=0.4\textwidth]{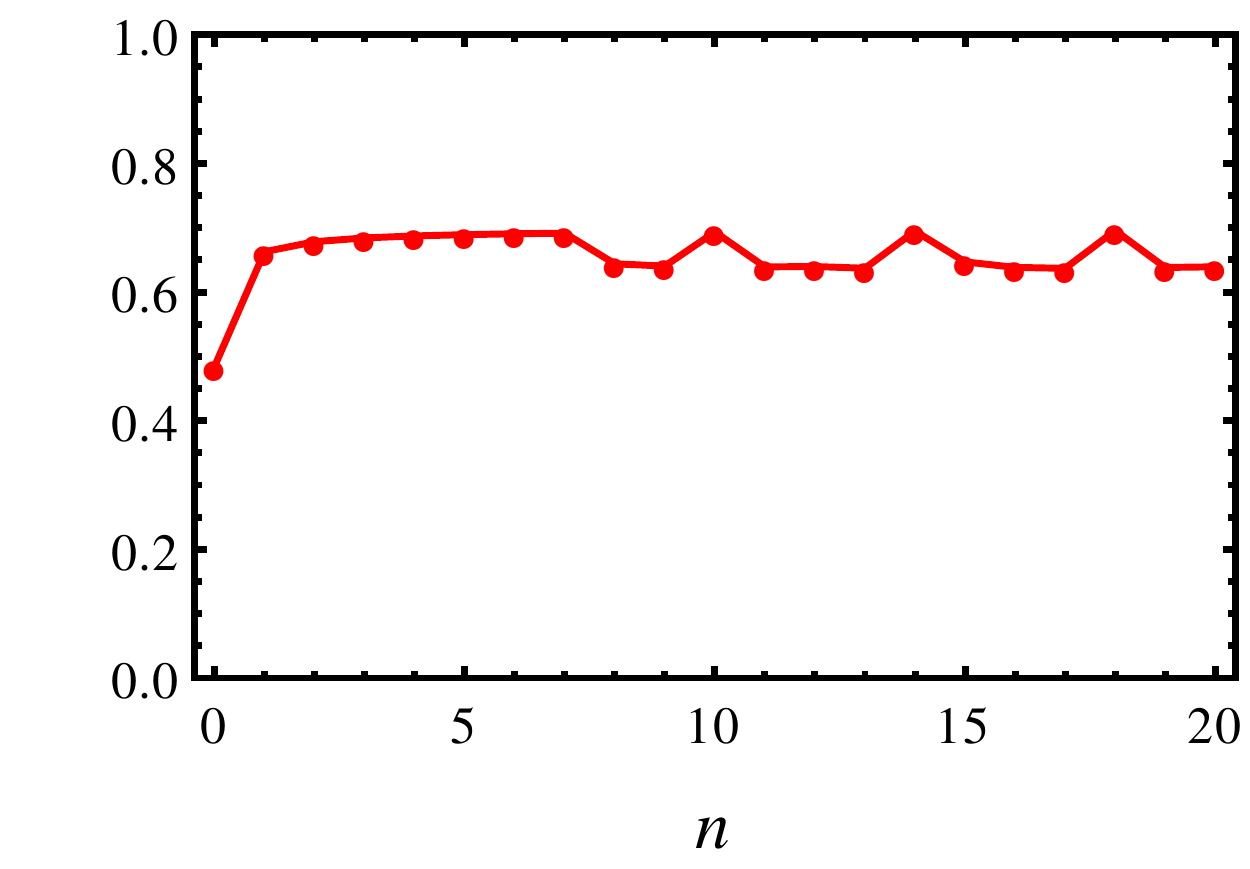} 
\caption{(Color online.) 
The reciprocity product in terms of Lipschitz constants for energy eigenstates of the quantum simple harmonic oscillator in one dimension. The horizontal axis represents the energy levels, while the vertical axis represents the reciprocity product, given in Eq.~(\ref{tilde_etax_etap}). The horizontal axis represents a dimensionless quantity, while the vertical one is in units of $1/\hbar$.} 
\label{excited_state_fig}
\end{figure}
\end{center}

In a similar fashion, we can find the reciprocity products for LCs of SHOs with $n=2,3,\ldots$.
We have performed the calculations to investigate the feature of the quantity $\tilde{\eta}_x \tilde{\eta}_p$ upto the $60^{\text{th}}$ excited level. Fig.~\ref{excited_state_fig} shows that among the energy eigenstates of SHO, the reciprocity product attains its lowest value for the ground state, and that after a few levels, the product has a value that is close to $0.63/\hbar$.
Therefore, for the quantum systems considered until now, we have
\begin{equation}
\tilde{\eta}_x \tilde{\eta}_p \geq \frac{1}{\hbar} \sqrt{\frac{2}{e \pi}}.
\label{gaussian_sqrt_etax_etap-natun}
\end{equation}

\begin{center}
\begin{figure}[t]
\includegraphics[width=0.4\textwidth]{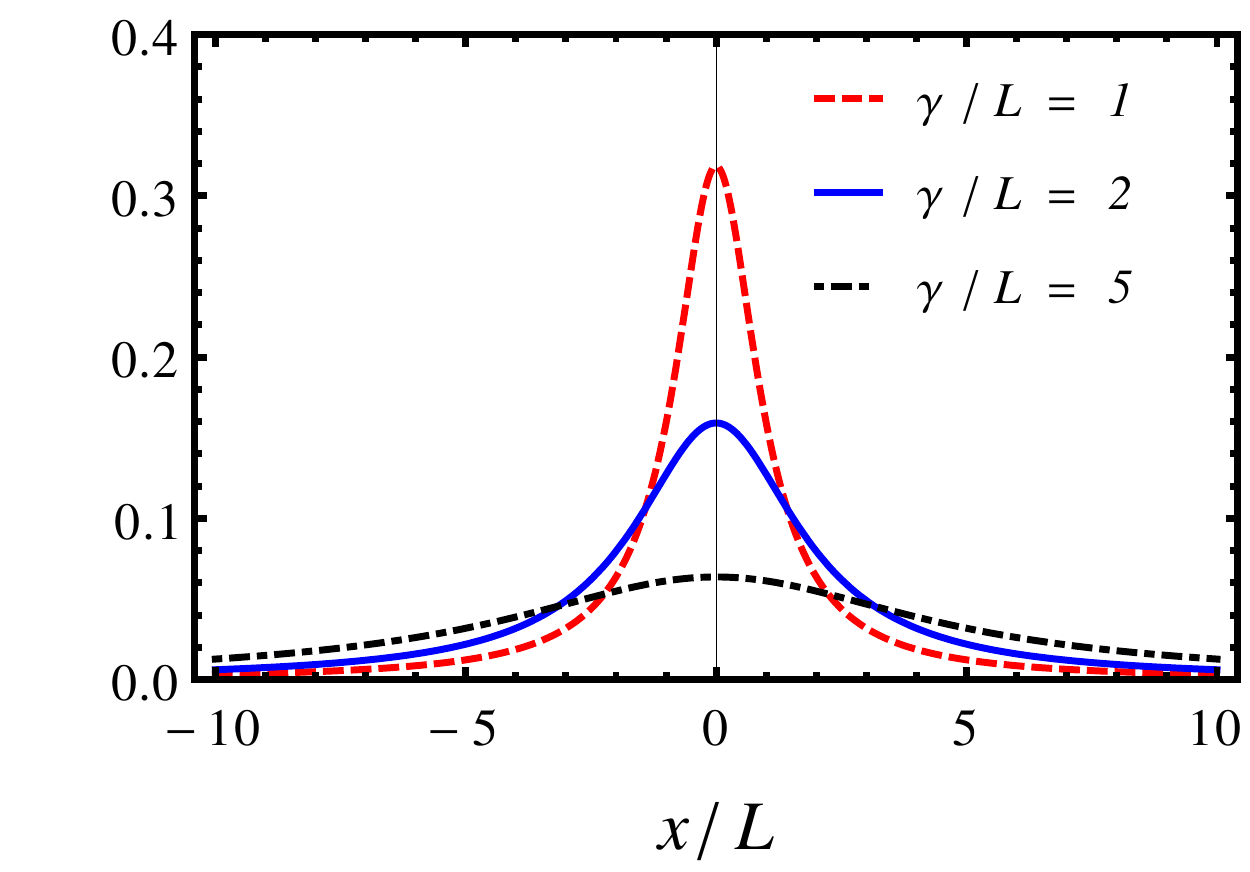} 
\caption{(Color online.) The Cauchy-Lorentz probability distribution. The function $Lf_c(x : x_{0} = 0, \gamma)$ is plotted on the vertical axis against $x/L$ for different values of $\gamma/L$. $L$ is a constant that has the unit of length. 
Both axes represent dimensionless quantities.} 
\label{chaucy_PDF_x}
\end{figure}
\end{center}

\section{Further probability distributions}
\label{other-prob}
Let us now consider quantum systems whose position distributions follow 
two other distributions, namely, 
the Cauchy-Lorentz  and the Student's $t$-distribution, to investigate the behavior of the quantity $\tilde{\eta}_{x}\tilde{\eta}_{p}$.

\subsection{Cauchy-Lorentz distribution}
\label{cauchy}
Let us now consider quantum systems whose position distributions follow 
the Cauchy-Lorentz distribution, introduced by M. G. Agnesi, S. D. Poisson, A.-L. Cauchy, H. A. Lorentz, and others. It is a continuous probability distribution function, given by
\begin{equation}
f_{c} (x : x_{0},\gamma) = \frac{\gamma}{\pi} \frac{1}{(x - x_{0})^{2} + \gamma^{2}},~x\in (-\infty,\infty), 
\end{equation}
where  $\gamma > 0$ and $x_{0}$ are distribution parameters. 
This corresponds to the position probability distribution of the state
 $\psi_{c} (x) = (f_{c} (x:x_{0},\gamma))^{\frac{1}{2}}, ~x\in (-\infty,\infty)$, of a quantum system moving on the $x$-axis. See Fig.~\ref{chaucy_PDF_x}.

Since the function $f_{c}(x)$ is differentiable,  one can calculate the Lipschitz constant in position space by maximizing the derivative of $f_c(x)$. We thereby obtain
\begin{equation}
\eta_{x} = \frac{3 \sqrt{3}}{8 \pi\gamma ^2}.
\end{equation}

\begin{center}
\begin{figure}[t]
\includegraphics[width=0.4\textwidth]{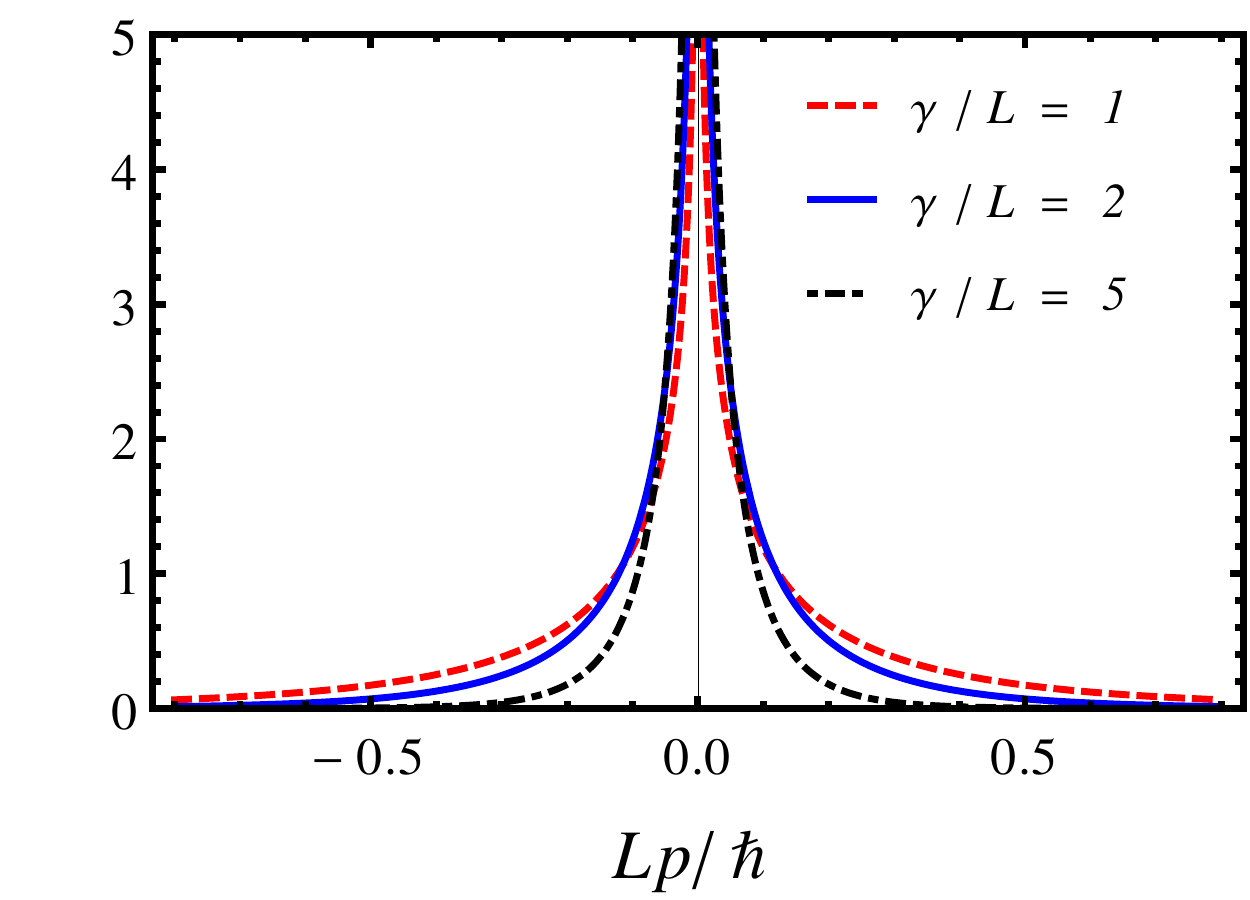} 
\caption{(Color online.)  Profile of $g_c(p)$, corresponding to the Cauchy-Lorentz probability distribution in position. The function $(\hbar/L)g_c(p:x_0=0,\gamma)$ is plotted on the vertical axis against $Lp/\hbar$ on the horizontal axis, for different values of $\gamma/L$.
The constant $L$ has unit of length. Both axes are dimensionless.} 
\label{cauchy_PDF_p}
\end{figure}
\end{center}

Now, in momentum representation, the wave function $\phi_{c}(p)$ corresponding to $\psi_c (x)$ is  given by
\begin{equation}
\phi_{c} (p) = \frac{\sqrt{\gamma}}{\sqrt{2 \pi^2 \hbar}} \int_{-\infty}^{\infty} \frac{e^{-ixp/\hbar}}{\sqrt{(x-x_0)^2+\gamma ^2}}dx.
\end{equation}
So, the corresponding momentum probability distribution is $g_c(p)=|\phi_{c} (p)|^2$. In Fig.~\ref{cauchy_PDF_p}, we plot this momentum probability distribution for different values of \(\gamma\).

\begin{center}
\begin{figure}[t]
\includegraphics[width=0.4\textwidth]{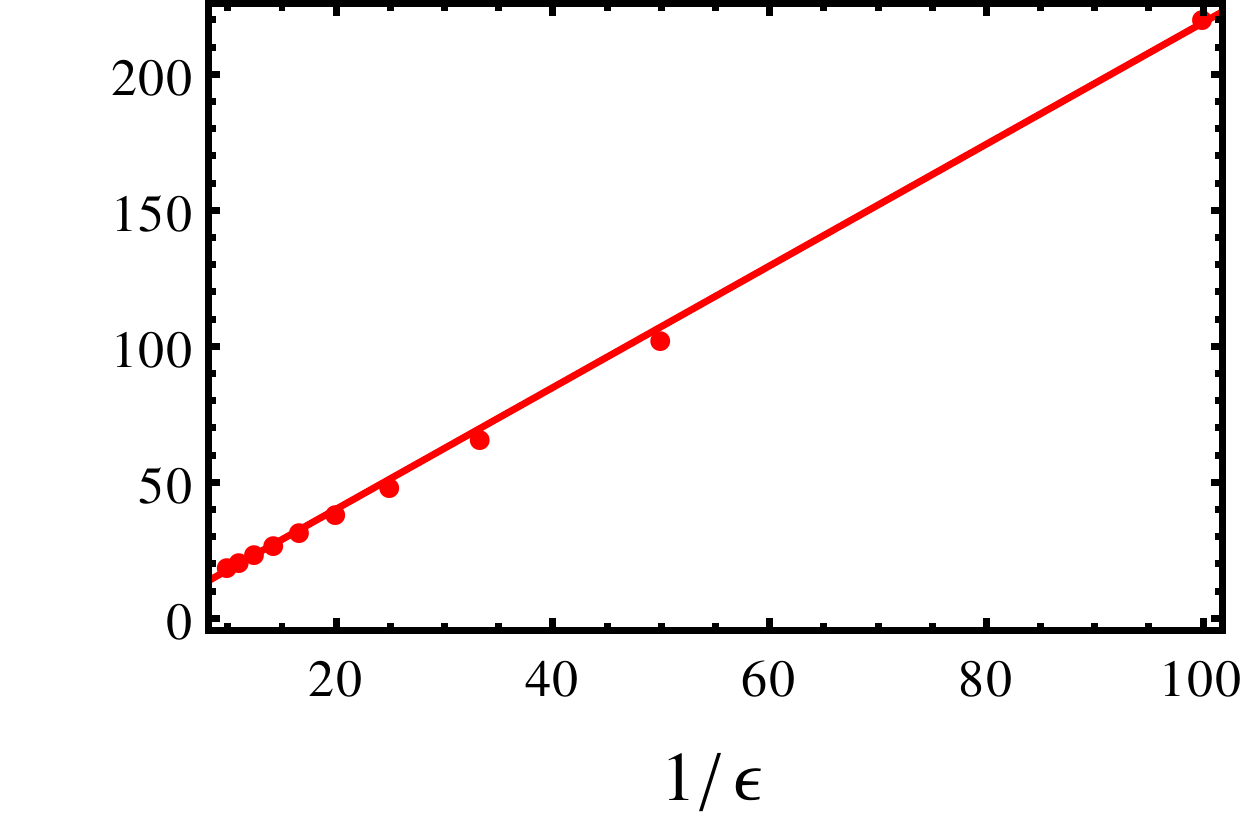} 
\caption{(Color online.) Divergence of the Lipschitz constant for $g_c(p)$. The value of the LC of $(\hbar/L)g_c(p:x_0=0,\gamma)$ considered as a function in the domain $\mathbb{R}_\epsilon$ is plotted on the vertical axis against $1/\epsilon$ on the horizontal axis. Both axes represent dimensionless quantities. We set $\gamma/L=1$. The red dots indicate the values of $1/\epsilon$ for which the LCs have been computed numerically. 
The dots are seen to fit to a straight line quite well.
Note that $\epsilon$ is a dimensionless 
quantity on the $Lp/\hbar$ axis.}
\label{cauchy_PDF_epsilon}
\end{figure}
\end{center}

The momentum distribution plots
suggest that the Lipschitz constant for
$g_c(p)$, denoted by $\eta_p$, diverges, irrespective of the values of $x_{0}$ and $\gamma$.
Moreover, the following argument will help us to understand this divergence more precisely. Consider the domain $\mathbb{R}_\epsilon$, where $\mathbb{R}_\epsilon=(-\infty, -\epsilon] \cup [\epsilon,\infty)$, $\epsilon$ being a small positive number. We have numerically estimated the value of the LC for the function $g_c(p)$ in the domain $\mathbb{R}_\epsilon$, for different values of $\epsilon$. The profile of the LC for $g_c(p)$ on $\mathbb{R}_\epsilon$ against $1/\epsilon$, as depicted in Fig.~\ref{cauchy_PDF_epsilon}, clearly indicates that $\eta_p$ diverges linearly with $1/\epsilon$. Therefore, we can say that $\tilde{\eta}_{x}\tilde{\eta}_{p}$ diverges for the quantum system corresponding to which the position distribution is Cauchy-Lorentz.

We therefore find that the $\tilde{\eta}_{x}\tilde{\eta}_{p} \to \infty$ point is shared by the classical scenario as well as systems represented by quantum states. We have already mentioned that 
$\tilde{\eta}_{x}\tilde{\eta}_{p} \to \infty$ for 
the classical case, viz. the product of two Dirac deltas in position and 
momentum. Here, we found that for the quantum state corresponding to which the position 
distribution is Cauchy-Lorentz, the reciprocity product $\tilde{\eta}_{x}\tilde{\eta}_{p}$ diverges. 
A similar feature will be seen for the case of the quantum system whose position distribution 
is the Student's \(t\) for two degrees of freedom.

The Cauchy-Lorentz distribution can be considered to be a constant function in the limit $\gamma \to \infty$. In this limit, $\eta_x$ converges to zero as $1/\gamma^2$. The momentum distribution has a diverging LC for all $\gamma$, including when $\gamma \to \infty$. This provides further evidence for a non-trivial lower bound of  $\tilde{\eta}_x \tilde{\eta}_{p}$, for all quantum states.
 
\subsection{Student's t-distribution}
\label{student}
\begin{center}
\begin{figure}[t]
\includegraphics[width=0.4\textwidth]{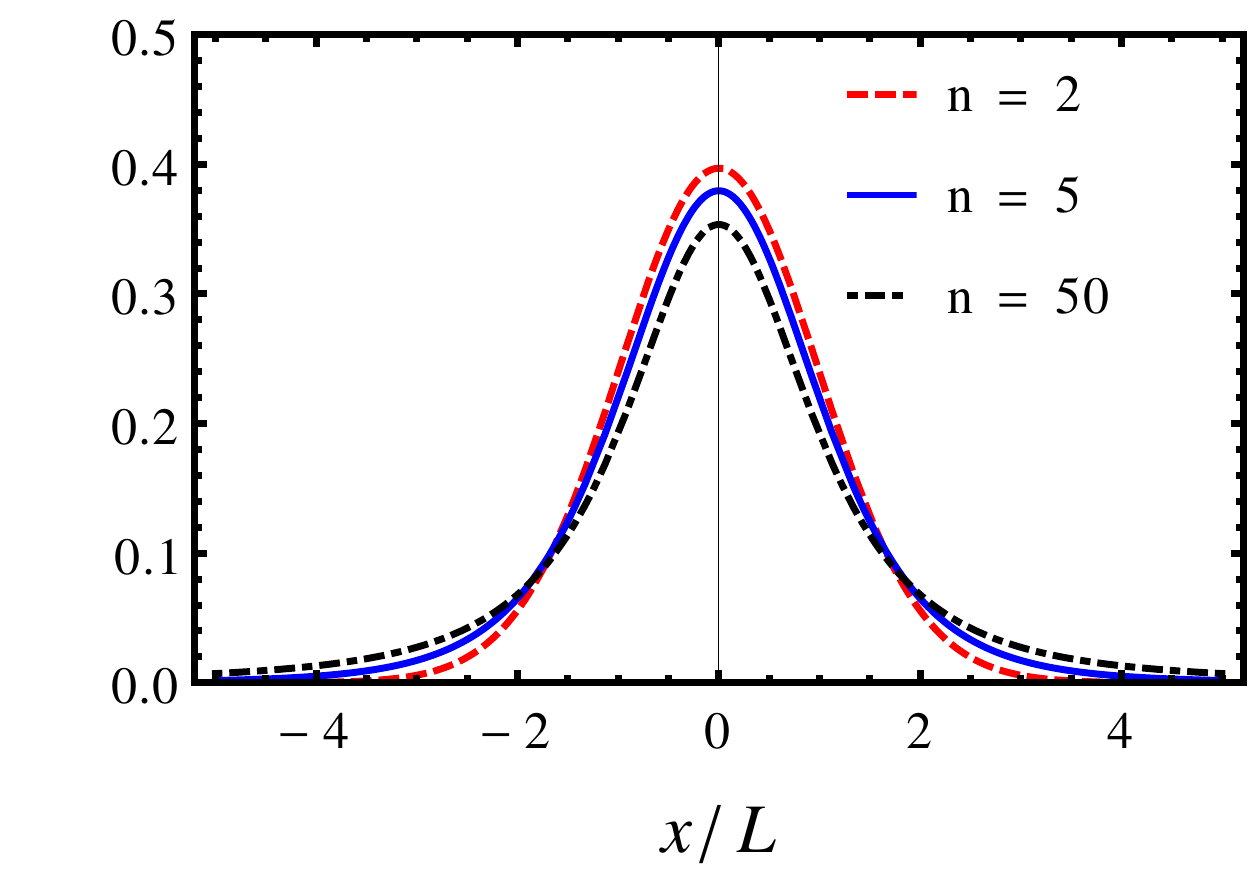} 
\caption{(Color online.) The Student's $t$ probability distribution function. We plot the function $L\tilde{f}_s(x,n:a,A)$, given by $\tilde{f}_s(x,n:a,A)=\frac{\Gamma(\frac{n + 1}{2})}{\sqrt{n \pi} \Gamma (\frac{n}{2})} A \left( a^2 + \frac{x^{2}}{n} \right)^{- \frac{n+ 1}{2}}$, against $x/L$, for $A/L^n=1$ and $a/L=1$. Both axes represent dimensionless quantities. As before, $L$ is a constant having the unit of length.}
\label{st_n_greater_2_x}
\end{figure}
\end{center}

Let us now consider the Student's $t$-distribution due to F. R. Helmert, J. L\"uroth, ``Student", and
others, for which the probability distribution function is given by
\begin{equation}
f_{s}(x:n) = \frac{\Gamma(\frac{n + 1}{2})}{\sqrt{n \pi} \Gamma (\frac{n}{2})} \left( 1 + \frac{x^{2}}{n} \right)^{- \frac{n+ 1}{2}}, ~x\in \mathbb{R}.
\end{equation}
Here $n$ is a distribution parameter, referred to as the number of degrees of freedom.
A graphical representation for $f_{s}(x:n)$ is given in Fig.~\ref{st_n_greater_2_x}.


It is interesting to start our discussion with two degrees of freedom, i.e., $n = 2$. Note that for Student's $t$-distribution with $n=2$, the mean exists but the standard deviation does not.

\begin{center}
\begin{figure}[h]
\includegraphics[width=0.4\textwidth]{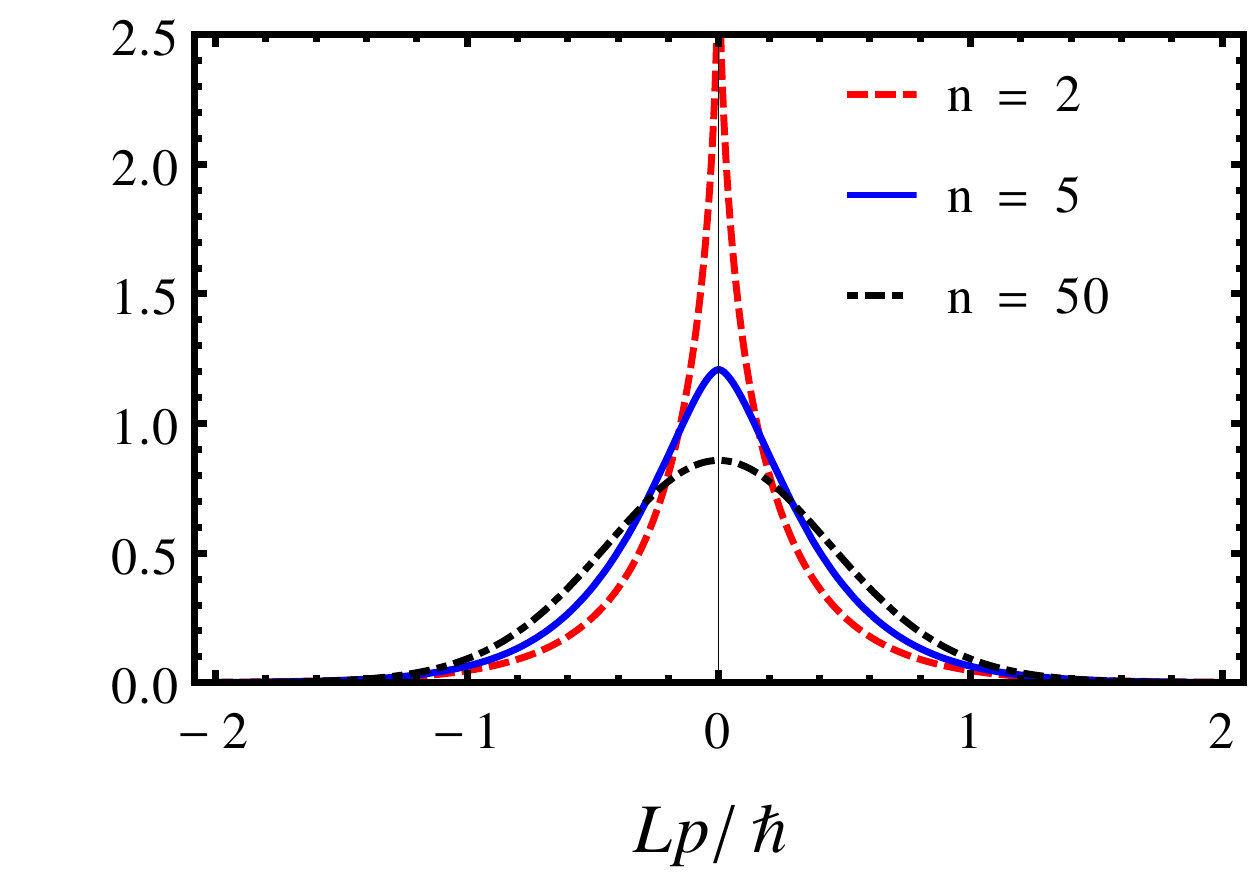} 
\caption{(Color online.) Profile of the momentum probability distribution corresponding to the Student's $t$-distribution. We plot $(\hbar/L)\tilde{g}_s(p, n: a, A)$, where $\tilde{g}_s$ is the Fourier transform of $\tilde{f}_s$, against $Lp/\hbar$. $L$ is again a constant with the unit of length, and we have chosen $A/L^n=1$ and $a/L=1$. Both axes are dimensionless.} 
\label{st_n_greater_2}
\end{figure}
\end{center}

Consider now a quantum system moving on the $x$-axis, and having the position distribution as $f_s(x:2)$.
Since the function $f_{s}(x:2)$ is  differentiable, the Lipschitz constant of the position distribution can be evaluated from the derivative of the distribution, and is given by
\begin{equation}
\eta_{x} = \frac{12}{25 \sqrt{5}} \approx 0.2147.
\end{equation}
In the momentum representation, by performing the Fourier transformation of $\psi_{s}(x:2)=(f_s(x:2))^{1/2}$, the momentum wave function is given by
\begin{equation}
\phi_{s}(p:2) = \frac{\sqrt[8]{2} \sqrt[4]{ \frac{\left|p\right|}{\hbar} } K_{-\frac{1}{4}}\left(\sqrt{2} \frac{\left|p\right|}{\hbar} \right)}{\sqrt{\hbar } ~ \Gamma \left(\frac{3}{4}\right)},
\end{equation}
where $K_{\nu}(z)$, $\nu \in \mathbb{R}$ is the modified Bessel function of the second kind.
The probability distribution in momentum space is $g_{s}(p:2)=|\phi_{s}(p:2)|^{2}$, as  depicted  in 
 Fig.~\ref{st_n_greater_2}, which suggests that the corresponding LC diverges.

\begin{center}
\begin{figure}[t]
\includegraphics[width=0.4\textwidth]{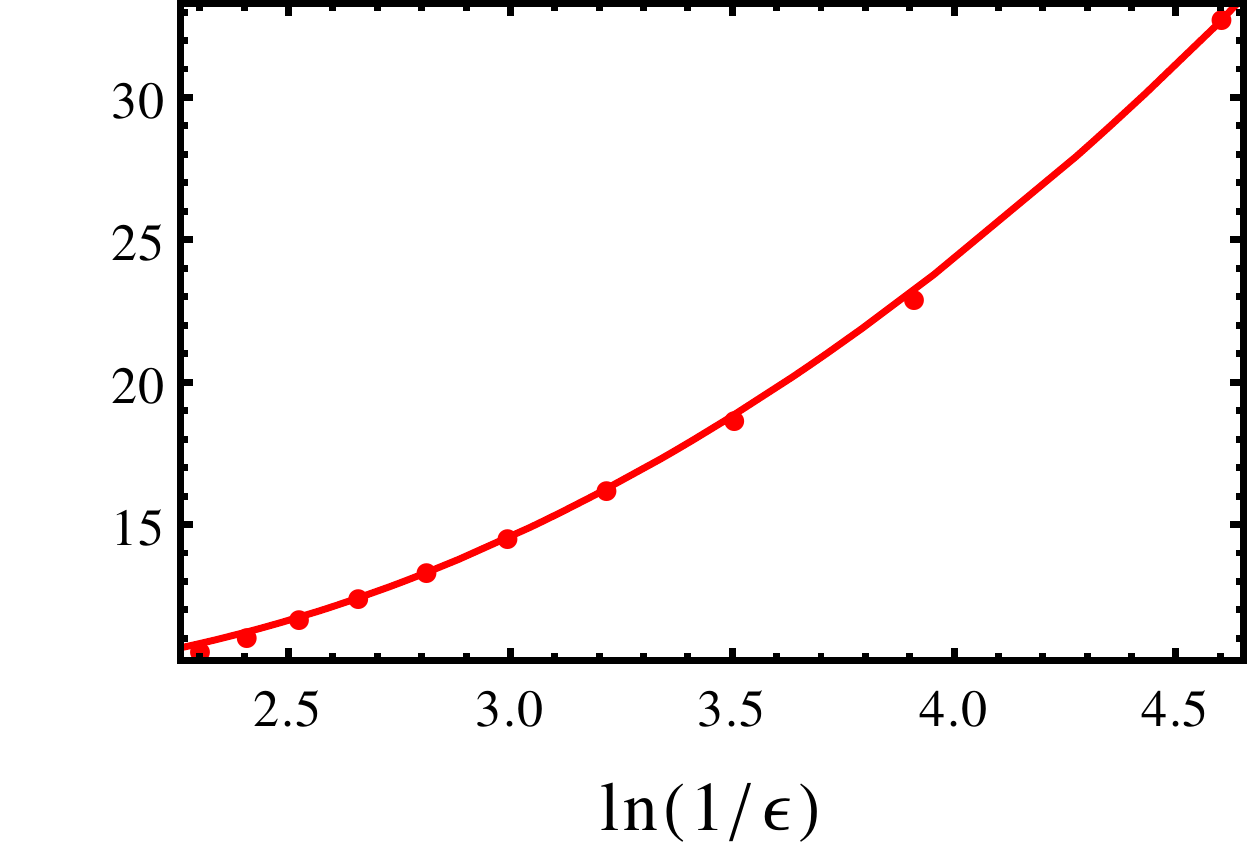} 
\caption{(Color online.) Behavior of the Lipschitz constant for the momentum distribution corresponding to the Student's $t$ distribution in position, for two degrees of freedom. We numerically compute the LC for $(\hbar/L) \tilde{g}_s$, considered as a function in the domain $\mathbb{R}_\epsilon$, for different values of $\epsilon$. The LC is then plotted on the vertical axis against $\ln(1/\epsilon)$ on the horizontal axis. Both axes represent dimensionless quantities. 
The red line is a quadratic fit of the data. It therefore follows that the LC diverges as \((\ln(1/\epsilon))^2\) for $\epsilon\to0$. The coefficient of the leading term in the quadratic divergence is $\approx 2.583$.}
\label{student_PDF_epsilon}
\end{figure}
\end{center}

The following argument makes this statement more precise. Consider the function $g_s(p:2)$ in the domain $\mathbb{R}_\epsilon$, as defined during our analysis of the Cauchy-Lorentz distribution. We find the LC for this function for varying values of $\epsilon$. This is depicted in Fig.~\ref{student_PDF_epsilon}, which clearly indicates a diverging LC for $\epsilon \to 0$. Consequently, the quantity  $\tilde{\eta}_{x}\tilde{\eta}_{p}$ diverges to infinity for $\psi_s(x:2)$.

\begin{center}
\begin{figure}[h]
\includegraphics[width=0.4\textwidth]{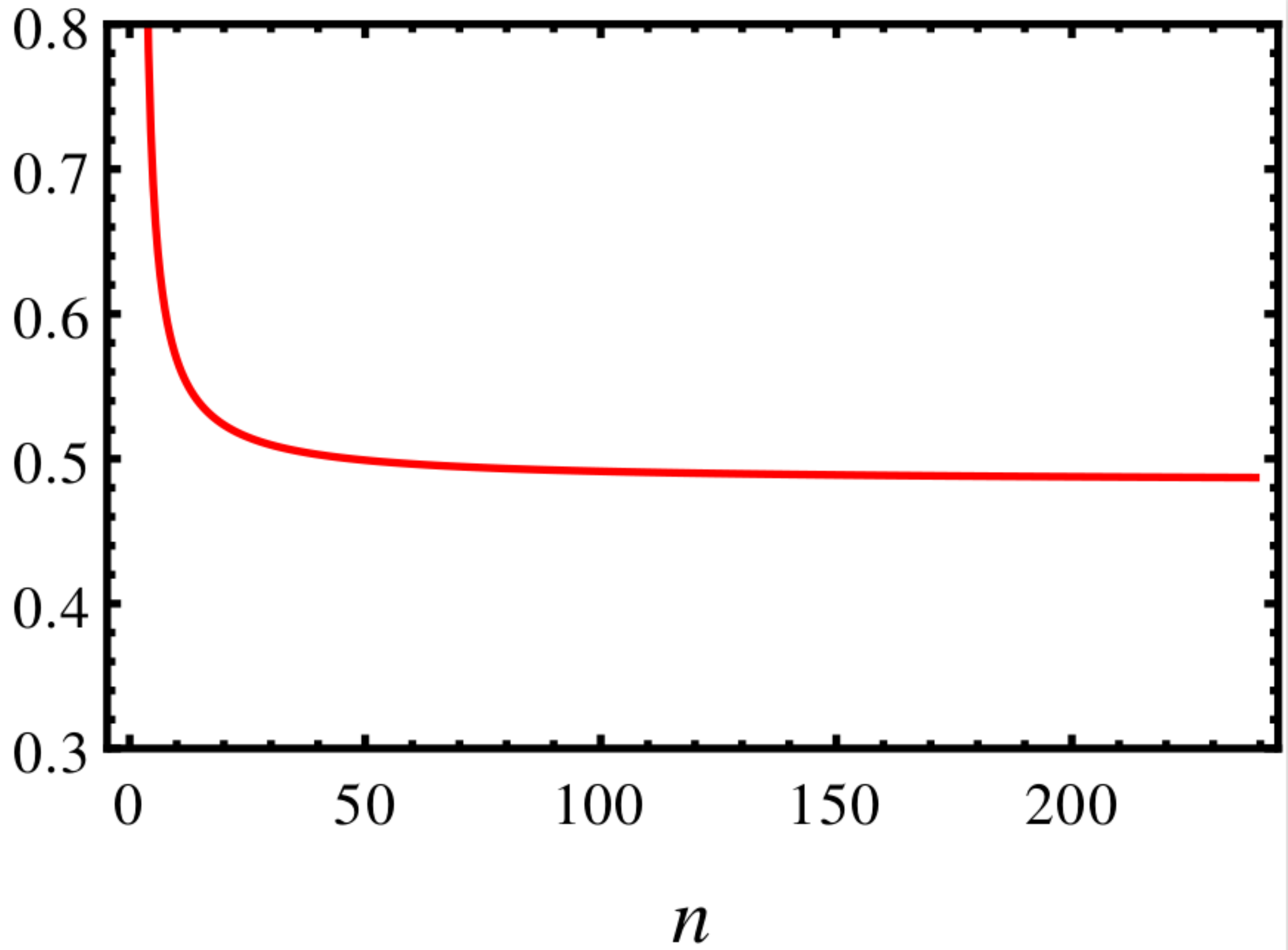}   
\caption{(Color online.) Reciprocity product of fluctuations for position and momentum distributions for the Student's $t$-distribution in position. The limit of large $n$ corresponds to the Gaussian distribution. Numerically, we obtain this value to be $\approx 0.48/\hbar$. This corresponds to the analytically obtained value of $(1/\hbar)\sqrt{2/{e\pi}}$. The horizontal axis is dimensionless, while the vertical axis is in units of $1/\hbar$.}
\label{st_n_greater_2_sqrt_etx_etap}
\end{figure}
\end{center}

The scenario changes for $n>2$. Indeed, for $n>2$, the momentum distribution provides a finite LC. It may be noted that the Student's $t$ distribution has finite variance for $n>2$, while the $n=2$ case does not provide a finite variance. 
It is known that for $n\rightarrow\infty$, the Student's $t$-distribution approaches to the Gaussian distribution, i.e., towards the ground state of simple harmonic oscillator. So it is expected that corresponding to the Student's $t$-distribution for $n\rightarrow\infty$, the quantity $\tilde{\eta}_{x}\tilde{\eta}_{p}$ will give the same value as that of the Gaussian distribution, as given in Eq.~(\ref{gaussian_sqrt_etax_etap}). 
 Fig.~\ref{st_n_greater_2_sqrt_etx_etap} shows that the reciprocity product in terms 
 of the LCs 
 of position and momentum distributions, corresponding to the Student's $t$-distribution in position, for increasing degrees of freedom, \(n\), converges to a limiting value of 
\(\approx 0.48/\hbar\), 
 just as for the Gaussian case (cf. Eq.~(\ref{gaussian_sqrt_etax_etap})).

\section{Completeness of polynomials and reciprocity relation} 
\label{hermite}
We now invoke the completeness of Hermite polynomials in $L_2 (- \infty, + \infty)$ \cite{simmonsbook} to determine the minimum-reciprocity states among all quantum states corresponding to systems of a single particle moving in one dimension. We Haar uniformly generate such functions, numerically, by considering polynomials until degree 5. The minimum-reciprocity product, $\tilde{\eta}_x \tilde{\eta}_p$, of the fluctuations of position and momentum in terms of the Lipschitz constants, for polynomials of different degrees, is given in  Table~\ref{table1}. To check for the efficiency of the numerical method, we have used it to find the minimum uncertainty bound in terms of variances. We numerically find that the minimum of the quantity $\Delta x \Delta p/\hbar$, by considering Hermite polynomials until degree 5, is 0.5, upto 6 decimal points, where $\Delta x$ and $\Delta p$ are the standard deviations in position and momentum respectively.
 
\begin{center}%
\begin{table}[h]
\begin{tabular}{|c|c|}
\hline
 $\quad n\quad $ & $\quad \hbar \tilde{\eta}_x \tilde{\eta}_p \quad $ 
\\ \hline
\hline 
  2  & 0.5     \\ \hline
 3 &  0.4        \\ \hline
 4 & 0.3   \\ \hline
 5 &  0.3   \\ \hline
\end{tabular}
\caption{
Numerical lower bound of quantum reciprocity product. We Haar uniformly generate \(N\) polynomials of degree \(n\), for \(n=2, 3, 4, 5\), where \(N\) is chosen to be sufficiently large so that convergence is obtained in the lower bound for a particular \(n\). 
We needed to have \(N = 3200, 12800, 25600, 102400\) respectively for \(n=2,3, 4, 5\).}
\label{table1}
\end{table}
\end{center}

\section{Interpreting the reciprocity relation and its differences with the uncertainty relation}
\label{Javed-Habib-BJP-te-join-korechhe}

The evidence presented in the preceding sections indicate that we have the relation
\begin{equation}
\tilde{\eta}_x \tilde{\eta}_p \gtrsim \frac{0.3}{\hbar}.
\end{equation}
In this section, we try to interpret this relation and also mention its differences with the uncertainty relation. 

Mathematically, the Lipschitz constant of a function gives an indication of how fast the function grows in its domain. For a given wave function, \(\psi(x)\), of a quantum particle moving in one dimension, \(\tilde{\eta}_x\) is the square root of the Lipschitz constant of the position probability distribution \(f(x) = |\psi(x)|^2\). The quantity, \(\tilde{\eta}_x\), of a wave function \(\psi(x)\) has been pressed into the job of quantifying the fluctuations in the position probability distribution \(f(x) = |\psi(x)|^2\). This is in contrast to the standard deviation, \(\Delta x\), of \(f(x)\), which quantifies the spread of the distribution \(f(x)\). Intuitively speaking, while \(\Delta x\) gauges the horizontal spread (i.e., along the abscissa) of the distribution \(f(x)\), \(\tilde{\eta}_x\) measures the vertical motions (i.e., along the ordinate) of the same. In this sense, the reciprocity relation uncovers an aspect of quantum states that is complementary to that revealed by the uncertainty relation. A parallel set of statements is true for the quantity, \(\tilde{\eta}_p\), of \(\phi(p)\), the Fourier transform of \(\psi(x)\), in relation to the standard deviation, \(\Delta p\), of the momentum probability distribution \(g(p)= |\psi(p)|^2\).

Just like the usual Heisenberg uncertainty relation lower bounds the product of the spreads of position and momentum distributions of a wave function, the reciprocity relation between \(\tilde{\eta}_x\) and \(\tilde{\eta}_p\) lower bounds the product of the fluctuations of position and momentum distributions of such a wave function. The Heisenberg uncertainty tells us that if the position distribution of a quantum particle is very well-defined, its momentum distribution must be very broad, and vice versa. Similarly, the reciprocity relation tells us that if the position fluctuations of a quantum particle are negligible, the momentum fluctuations must be very large. 

The Heisenberg uncertainty relation between two observables emanates from the distinctly quantum property of noncommutativity of the corresponding observables. We note in particular that \( \left[ x,p \right] = i \hbar \). It is plausible that the reciprocity relation between any two observables is also linked to a quantum property of the observables. 

The Heisenberg uncertainty relation between position and momentum is known to be saturated for Gaussian probability distributions in position. Our numerical simulations imply that the reciprocity relation between position and momentum is saturated by certain Hermite polynomials of degree 4 and 5.

It is in order here to present a few words about the classical limit of this inequality. Traditionally, it has been argued that the Heisenberg uncertainty limit and other relations in quantum mechanics converge to their classical cousins in the limit of \(\hbar \to 0\). This approach however has several difficulties, as mentioned, for example, in \cite{Ballentine-paper-1994,sen-ehrenfest,classicallimit,Angelo}. In spite of these difficulties, one usually claims that the quantum mechanically allowed region for the uncertainty product of spreads for position and momentum, stretches from a factor of the Planck's constant till the positive infinity, while 
the classical case (for pure states) is situated at zero. The classical case in this scenario is 
a product of two Dirac deltas, respectively in position and momentum.
There is a region, on the axis of the uncertainty product of spreads, near zero, that is forbidden to all quantum states, so that there is a gap between the quantum mechanically accessible region and 
the accessible point for the classical case. The reciprocity product of fluctuations, again has a quantum mechanically forbidden region adjoining  zero. However, the length of the quantum mechanically forbidden region, on the axis of the reciprocity product of fluctuations, is of the order of  inverse of the Planck's constant. Moreover, the classical case is at one end of the quantum mechanically accessible region, instead of being in the quantum mechanically forbidden region. This is akin to the case where we have an ensemble of two non-orthogonal pure quantum states~\cite{Christopher}, with the ensemble being a function of the angle between the two state vectors. The limit of an ensemble of two orthogonal states -- the classical case -- is situated at one end but within the quantum mechanically accessible region, on the axis of the angle between the two vectors.

\section{Conclusion}
\label{conclude}
The position and momentum probability distributions of arbitrary quantum
states are constrained by the Heisenberg uncertainty relation. In
particular, their spreads cannot both be arbitrarily small. We proposed an
independent restriction of the same distributions for arbitrary quantum
states, and termed it as the quantum reciprocity relation. 
We found that Lipschitz constants of the position and momentum distributions cannot both be arbitrarily
small. The lowest value of the product of the square roots of the Lipschitz
constants was found by invoking the completeness of the Hermite polynomials in the space of square-integrable functions.
Specific quantum states that were considered in the analysis include
the ground and excited states of the quantum simple harmonic oscillator, and quantum
states corresponding to which the position distribution is Cauchy-Lorentz or
Student's $t$ distributions. 
It is to be noted that the proposed bound has not yet been proven analytically, and is  currently based on numerics and specific
examples.

\begin{acknowledgments}
A.B. acknowledges the support of the Department of Science and Technology (DST), Government of India, through the award of an INSPIRE fellowship.
\end{acknowledgments}



\pagebreak

\textbf{Reply to ``Comment on `Quantum reciprocity relations for fluctuations of position and momentum'''}

\textbf{Abstract.}
\emph{In the article above, we had proposed an inequality concerning Lipshitz constants of position and momentum distributions, based on some numerical evidence and some examples. In a comment on it, this has been proven to be incorrect, by providing an example that violates the inequality. In this reply, we wish to state that the original article was based on a certain intuitive belief about fluctuations of position and momentum wave functions of a quantum state, which itself may still be correct. The way in which we tried to enunciate its mathematical form has of course been proven to be wrong. We indicate, in this reply, a possible alternate mathematical form that such a relation may assume.}

We thank the author of the comment [I. Bia{\l}ynicki-Birula, Phys. Rev. A \textbf{100}, 046101 (2019)] for the example. We wish
to add the following.

First of all, we wish to mention that our proposed inequality is not an
uncertainty for position and momentum, as Lipshitz constants do not
quantify the spread of a function. While an uncertainty relation concerns
spreads along the ``horizontal'' axes (abscissae) of the relevant
functions, our proposed relation (``reciprocity relation'') concerns
fluctuation of the function values themselves (``vertical'' axes, i.e.,
ordinates).

The comment clearly provides a counterexample to our proposed inequality
in terms of the Lipshitz constants of the position and momentum
distributions of quantum wave functions. However, we wish to note that the
point of departure for us was that if the position axes are
``featureless'', the momentum axes cannot be (and vice versa). The
counterexample uses a wave function that deftly ``stores'' the position
features in a phase. Eq. (1) of the comment is a wave function in position
that has a lot of features even if \(a\) is very large. These features are
stored in the parameter \(b\), which is finally made very large (after Eq.
(12)). In absence of \(b\), Eq. (1) is featureless for a large \(a\). But
not so, if \(b\) is present, and especially if \(b\) is large. These
features of the wave function in Eq. (1), due to the presence of \(b\),
are washed out in the position \emph{probabilities}, but are present in
the position wave function.

A quantum system moving in one space dimension (\(x\)) and that is
featureless in position is to weakly depend on \(x\), even in its phase.
We agree that we did not put this constraint in our paper, instead
requiring only that the position \emph{probability} be weakly dependent on
\(x\).

We believe that there will exist a nontrivial lower bound for the product
of the Lipshitz constants of position and momentum distributions of
arbitrary quantum mechanical wave functions in position that satisfies the
condition of being weakly dependent on position on all axes on the complex
plane on which the wave function is defined. The phrase “weakly dependent''
could, e.g., be defined again by employing the Lipshitz constant along the
corresponding axis of the said complex plane.

For a general quantum mechanical wave function that does not satisfy the
weak dependence criterion, one needs to identify a quantity that involves
Lipshitz constants along all axes (or possibly, just the real and
imaginary axes) for the wave function in coordinate and momentum
representations, instead of simply the product of Lipshitz constants of
position and momentum probability distributions.

We thank Michael Hall for useful comments.

\end{document}